\documentclass[12pt]{iopart}

\usepackage{graphicx}% Include figure files
\usepackage{dcolumn}% Align table columns on decimal point
\usepackage{bm}% bold math
\usepackage{amssymb,amsthm}   % for math
\usepackage{color}		 % for editing purposes
\usepackage{epstopdf}	 % .eps figures
\usepackage{cases}
\usepackage{iopams}
\usepackage{mathrsfs}
\usepackage{amsfonts}
\usepackage{fancyhdr}
\usepackage{enumerate}
\usepackage{multirow}
\begin{document}

\title[]{Self-consistent simulations of ECR-based charge breeders: evidence and impact of the plasmoid-halo structure}

\author{A. Galat\`a$^1$, C. S. Gallo$^{1,2}$, D. Mascali$^{3}$, G. Torrisi$^{3}$ and M. Caldara$^{4}$}

\address{$^1$INFN-Laboratori Nazionali di Legnaro, Legnaro, Padova, Italy}

\address{$^2$Dipartimento di Fisica e Scienze della Terra, Universit\`a di Ferrara, Ferrara, Italy}

\address{$^3$INFN-Laboratori Nazionali del Sud, Catania, Italy}

\address{$^4$Dipartimento di Fisica e Astronomia, Universit\`a di Padova, Padova, Italy}

\begin{abstract}
This paper discusses the capture of an ion beam in a magnetized plasma of an Electron Cyclotron Resonance Ion Source based Charge Breeder, as modelled by numerical simulations. As a relevant step forward with respect to previous works, here the capture is modeled by considering a plasma structure determined in a self-consisent way. The plasmoid-halo structure of the ECR plasma - that is consisting of a dense core (the plasmoid) surrounded by a rarefied halo - is further confirmed by the self-consistent calculations, having also some fine structures affected by the electromagnetic field distribution and by the magnetostatic field profile. The capture of Rb$^{1+}$ ions has been investigated in details, vs. various plasma parameters, and then compared to experimental results.
\end{abstract}

%\noindent{\it magnetoplasmas, numerical simulations, ion sources, charge breeders\/}

%\submitto{\PSST}

\maketitle

%%%%%%%%%%%%%%%%%%%%%%%%%%%%%%%%%%%%%%%%%%%%%%%%%%%%%%%%%%%%%%%%%%%%%%%
\section{Introduction}

For the last years the INFN ion source group has been dedicating its efforts to the development of a predictive numerical tool for magnetically confined plasmas~\cite{EPJ}, in particular for Electron Cyclotron Resonance (ECR) Ion Sources~\cite{ECR} or Traps and ECR-based Charge Breeders. The aim is to determine, for any given excitation frequency, the spatial density and energy distributions for both electrons and ions. Presently, the work is being carried out in the framework of the PANDORA~\cite{Pandora} (Plasmas for Astrophysics, Nuclear Decays Observation and Radiation for Archeometry) project, in the past financed by the 5$^{th}$ and now by the 3$^{rd}$ Scientific Committe of INFN. This innovative facility, based on a magnetic plasma trap, will be used to carry out fundamental studies in nuclear astrophysics, astrophysics and plasma physics, by measuring nuclear decay rates in stellar like conditions as a function of the ionization state. The PANDORA plasma will be based on the so called ``B-minimum'' magnetic structure, typical of ECR ion sources: low activity radioactive species will be injected by employing conventional techniques used by the ion source community (gas bottles, resistive ovens, sputtering systems); for the high activity ones, the charge breeding technique will be employed, widely used in Isotope Separation On Line (ISOL) facilities whose aim is the post-acceleration of radioactive ions for nuclear physics experiments, as in the case of the SPES project~\cite{SPES}. A complete plasma diagnostic set-up is under development: nuclear decays will be tagged by the gamma emission of excited nuclei or, when possible, by optical emission spectroscopy of the decay products. Spatially resolved X-ray measurement will give information on ions and electrons density, as well as electrons temperature: this work is presently carried out in collaboration with the Hungarian laboratory ATOMKI~\cite{X-ray ATOMKI}. Finally, the charge state distribution will be measured by optical emission spectroscopy using the high resolution spectrograph SARG, made available to the PANDORA collaboration by the National Institute of Astrophysics. In view of an efficient spatially resolved diagnostics within PANDORA, the knowledge of the plasma fine structure is mandatory: to this scope, strong efforts have been making to obtain a self-consistent ECR-plasma description, by joining precise electromagnetic calculations, carried out with COMSOL-Multiphysics, with the electrons dynamics calculated with MatLab. In this paper the authors present the results of self-consistent simulations of magnetically confined plasmas typical of ECR ion sources and charge breeders: the plasmoid/halo stucture~\cite{Wiesemann}, that is a dense plasma core surrounded by a rarefied halo, has been demonstrated numerically for the first time. The results highlight also an internal fine structure of the plasmoid, evidencing the possibility of a ``hallow'' shape of the ECR plasmas. Finally, the obtained plasma target model has been used to simulate the charge breeding of Rb$^{1+}$ ions and compare the results with those presented in~\cite{PSST_CB}, evidencing the influence of the various plasma parameters on the capture process.
%%%%%%%%%%%%%%%%%%%%%%%%%%%%%%%%%%%%%%%%%%%%%%%%%%%%%%%%%%%%%%%%%%%%%%%%%%%%
\section{The self-consistent approach}
\label{sec:2}

\begin{figure}%[ht]
\centering
\includegraphics[width=1\textwidth]{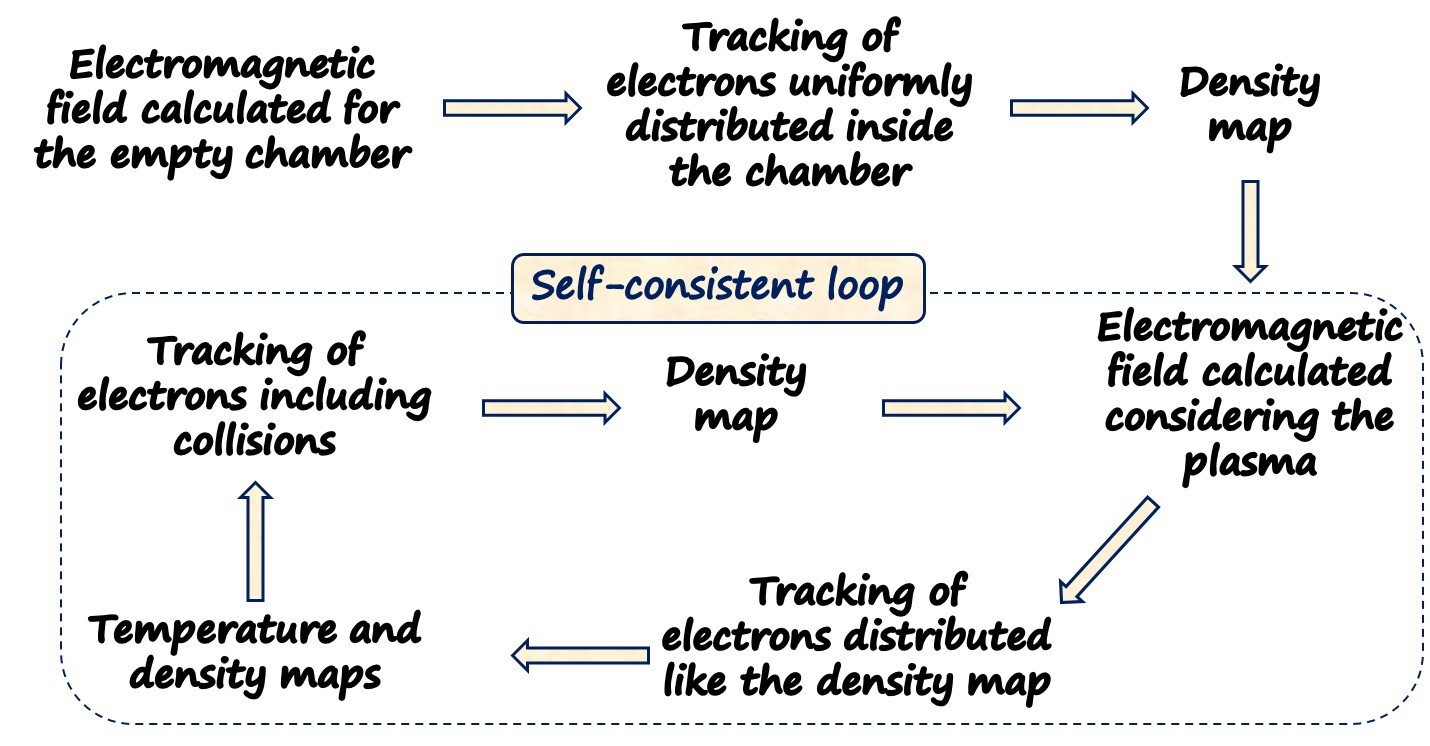}
\caption{Steps executed by the Comsol and MatLab codes to reach self-consistency for electrons dynamics.}
\label{fig:1}
\end{figure}

The self-consistent modelling is achieved by an iterative procedure that aims at solving the collisional Vlasov-Boltzman equation; details are discussed in~\cite{EPJ}. By this approach, both an electromagnetic solver such as COMSOL-Multiphysics and a kinetic code (written in MatLab) for solving particles' equation of motion are used in an iterative process, assuming a stationary plasma. The two approaches must be necessarily used at the same time: in fact, on one hand in magnetized plasmas excited by microwaves (as is the case of ECR sources, traps and charge breeders) the electromagnetic field set-up inside the plasma chamber determines, through a resonant interaction, the energetic content of electrons and then the plasma density. On the other hand, the plasma is an anisotropic and dispersive medium characterized by a 3D dielectric tensor, that must be taken into account for the calculation of the electromagnetic field in a kind of self-consistent loop~\cite{Turco}. In order to reach such a self-consistency we followed a step-wise approach:
\begin{enumerate}
\item Simulations start with the calculation of the electromagnetic field in an empty plasma chamber. Then, this field is used to calculate the dynamics of electrons generated uniformly inside the plasma chamber, by using a MatLab code derived directly from the one developed for ions and described in~\cite{PSST_CB}, and including the confining magnetostatic field and relativistic effects. The code follows the evolution of N electrons for a fixed simulation time T$_{span}$: at this step particles dynamics is determined only by electromagnetic and magnetostatic fields. By using an ad-hoc routine, the code stores particles' positions at each time step in a 3D matrix reflecting the domain of the simulation divided into cells of 1 mm$^3$, creating an ``occupation'' map. The occupation map obtained after this first ``vacuum'' step is then scaled to a density map by supposing an equivalent total number of particles compared to the one obtained by the 3D plasmoid/halo density map used in~\cite{PSST_CB}. In that case we had $n_{plamoid}\sim$ 2$\cdot$10$^{17}$ m$^{-3}$ and $n_{halo}$= $n_{plamoid}$/100. This calculation concludes what we consider the Step-0.
\item The density map obtained after the Step-0 is used to calculate the value of the plasma dielectric constant of the magnetized plasma in each cell: the Step-1 starts with new electromagnetic calculations, this time including a plasma (through its dielectric constant) distributed like the previously calculated density map. The new electromagnetic field is used to calculate again electrons dynamics including the magnetostatic field, creating this time not only an occupation (and so a density) map but also an energy map, obtained storing at each time step the energy of the electrons in the elements of another 3D matrix, corresponding to their spatial location. By dividing the energy map to the occupation map, the spatial distribution of the average energy is obtained: by supposing it belongs to electrons with a Maxwell-Boltzmann distribution, a spatial temperature map can be derived by multiplying the average energy for the factor (3$\pi$/8).
\item Electrons dynamics is calculated again including this time the presence of the plasma, whose temperature and density are locally determined by the relative maps and used to calculate the frequency of elastic electron-electron collisions~\cite{Spitzer}. The output of this calculation concludes the Step-1, with a new density map that can be used to calculate again the 3D dielectric tensor and then the electromagnetic field.
\end{enumerate}
Step-1 is repeated until the results show self-consistency, that is until the results from consecutive steps show negligible differences in the density and temperature matrices. The scheme shown in figure~\ref{fig:1} explains the various steps of the self-consistent loop.
\section{Simulation results}
\begin{figure}%[ht]
\centering
\includegraphics[width=1\textwidth]{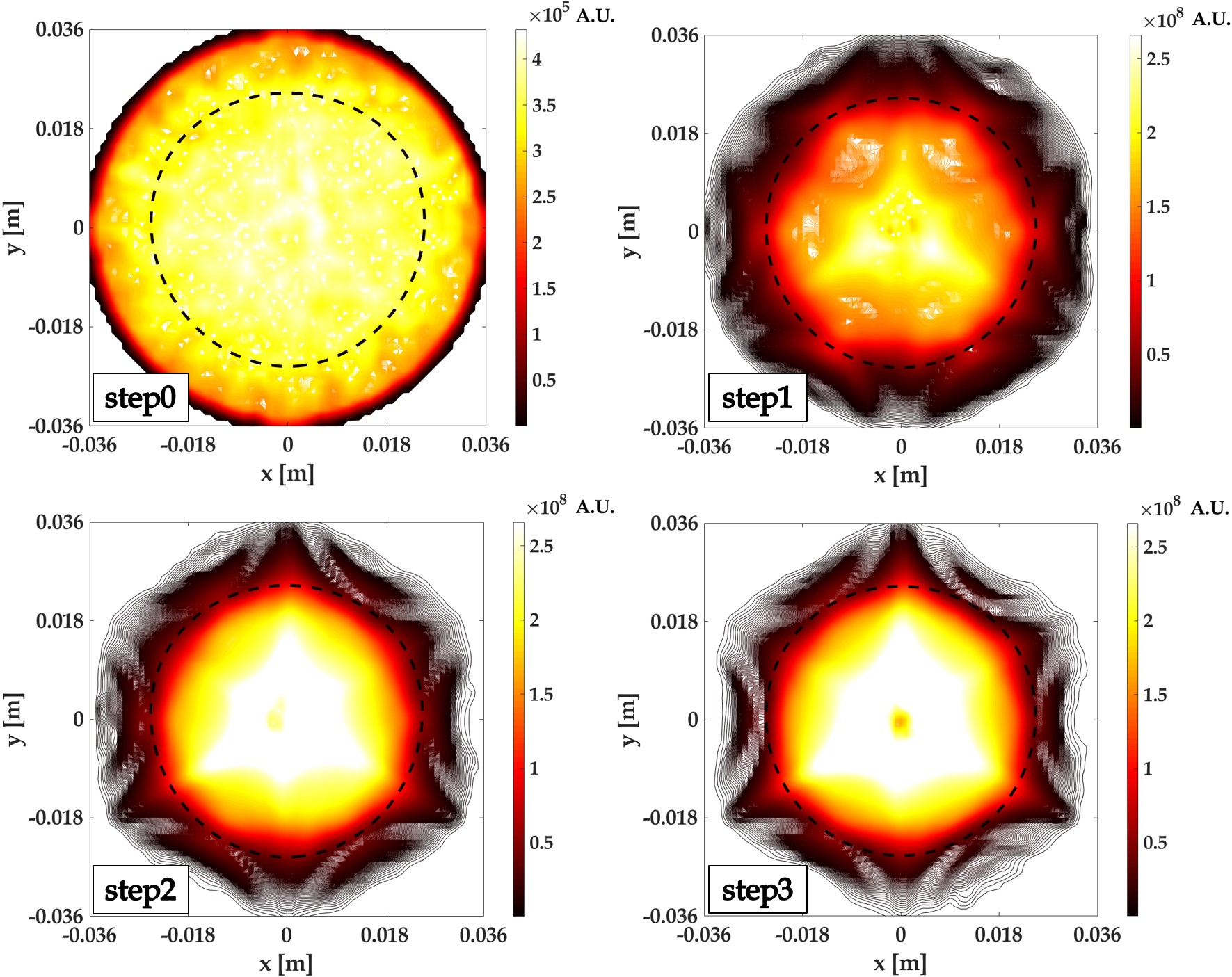}
\caption{Evolution of the projection along the plasma chamber axis of the occupation maps during the calculation steps of the self-consistent loop. The black dotted line indicates the contour of the resonance surface on the midplane. }
\label{fig:2}
\end{figure}

\begin{figure}%[ht]
\centering
\includegraphics[width=0.8\textwidth]{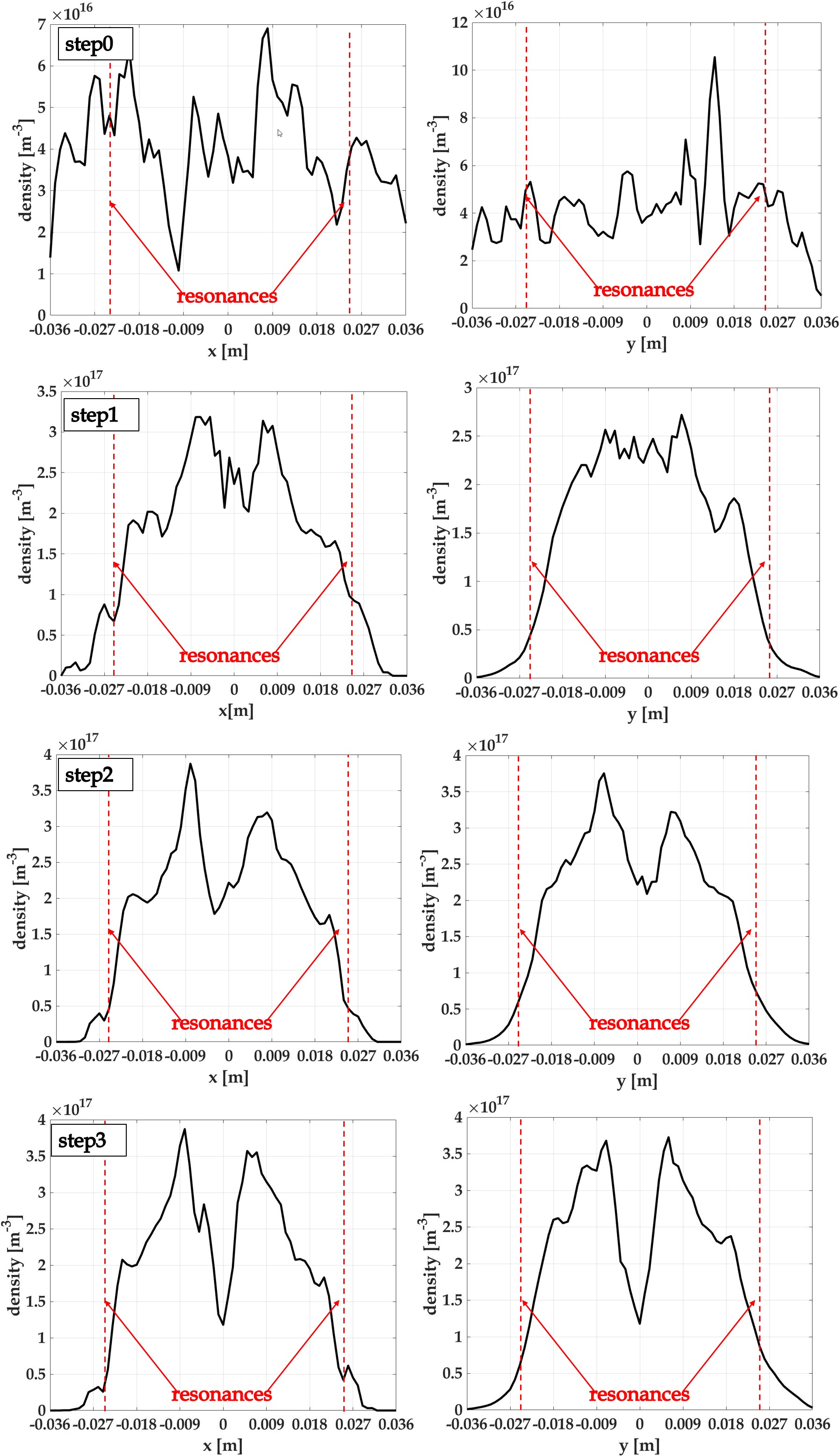}
\caption{Evolution of the density profiles on the midplane during the calculations of the self-consistent loop: x direction (left), y direction (right). The red lines indicate the location of the resonance zones. }
\label{fig:3}
\end{figure}

The geometry employed for the electromagnetic simulations is the stainless steel plasma chamber of the SPES project charge breeder~\cite{SPES_CB}: it is a cylinder 353 mm long and 36 mm in radius. Its boundaries are an aluminum plate at injection side, with a r$_{1+}$ = 14 mm hole to allow the 1+ beam injection, and the r$_{ext}$ = 4 mm extraction electrode (extraction side). COMSOL-Multiphysics implements an adaptative mesh refinement to calculate the electromagnetic field, with a maximum and minimum spacing respectively of $\lambda$/3.5 and $\lambda$/35, with $\lambda$ the wavelength corresponding to an operating frequency of 14.521 GHz. Microwaves are radially injected through a rectangular WR-62 waveguide located 10 mm from the injection side, with a power of 100 W. Previous results of this approach with the step-wise plasmoid/halo structure used in~\cite{PSST_CB} can be found in~\cite{CB_EM_SIM}. In this region, a minimum-B geometry is created by three coils and a permanent magnet hexapole: the maximum at injection is B$_{inj}$ = 1.18 T, located at 66 mm from the injection side, at extraction and on the plasma chamber walls B$_{ext}$ = B$_{rad }$= 0.79 T, while the minimum is B$_{min}$ = 0.35 T. The distance between the two maxima is 288 mm: this space is the simulation domain for electrons dynamics, where the code developed in MatLab integrated the equation of motion of N=40000 electrons, with a time step T$_{step}$=10$^{-12}$ s and a total simulation time T$_{span}$=40 $\mu$s. For this specific frequency, self-consistency has been obtained after three iterations of the loop: figure~\ref{fig:2} shows the evolution of the occupation maps, projected along the plasma chamber axis. It is interesting to see how the density, starting from a uniform distribution, concentrates progressively inside the resonance surface (whose contour in the midplane is indicated by the black dotted line in the figure), thus forming the plasmoid. It, in turn, is far from being uniform and shows the typical ``star shape'' impressed to the plasma by the hexapolar field.
A hollow nature of the plasmoid is clearly seen in figure~\ref{fig:3}, showing the evolution of the density profiles along the x and y axes on the midplane, after conversion from the occupation map; the vertical red lines indicate the points corresponding to the resonances. Starting from a condition in which the plasma is distributed inside and outside the resonance, a dense plasmoid is progressively formed, acquiring finally a hollow shape. It is important to note that, inside the plasmoid, the density reaches a value $n_{max}$ almost twice the one deduced considering a uniform plasmoid~\cite{PSST_CB} ($n_{max}\sim$ 4$\cdot$10$^{17}$ m$^{-3}$) in two points off axis, and descreases of a factor of 3 going towards the center and 7 going toward the resonance surface. Just outside the resonance, the density is higher than the one considered in~\cite{PSST_CB} for the halo, but then decreases very fastly towards the plasma chamber walls: this behaviour is quite similar to the one observed for the profiles of X-rays emitted by the plasma of the ECR source installed at ATOMKI (both spectrally integrated and filtered for the Ar K-alpha peak), as described in~\cite{X-ray ATOMKI}. Compared to the calculation presented in~\cite{EPJ}, this can be considered a confirmation that the plasmoid/halo structure is automatically formed by the wave-plasma interaction for any initial distribution of plasma electrons.

\begin{figure}%[ht]
\centering
\includegraphics[width=0.95\textwidth]{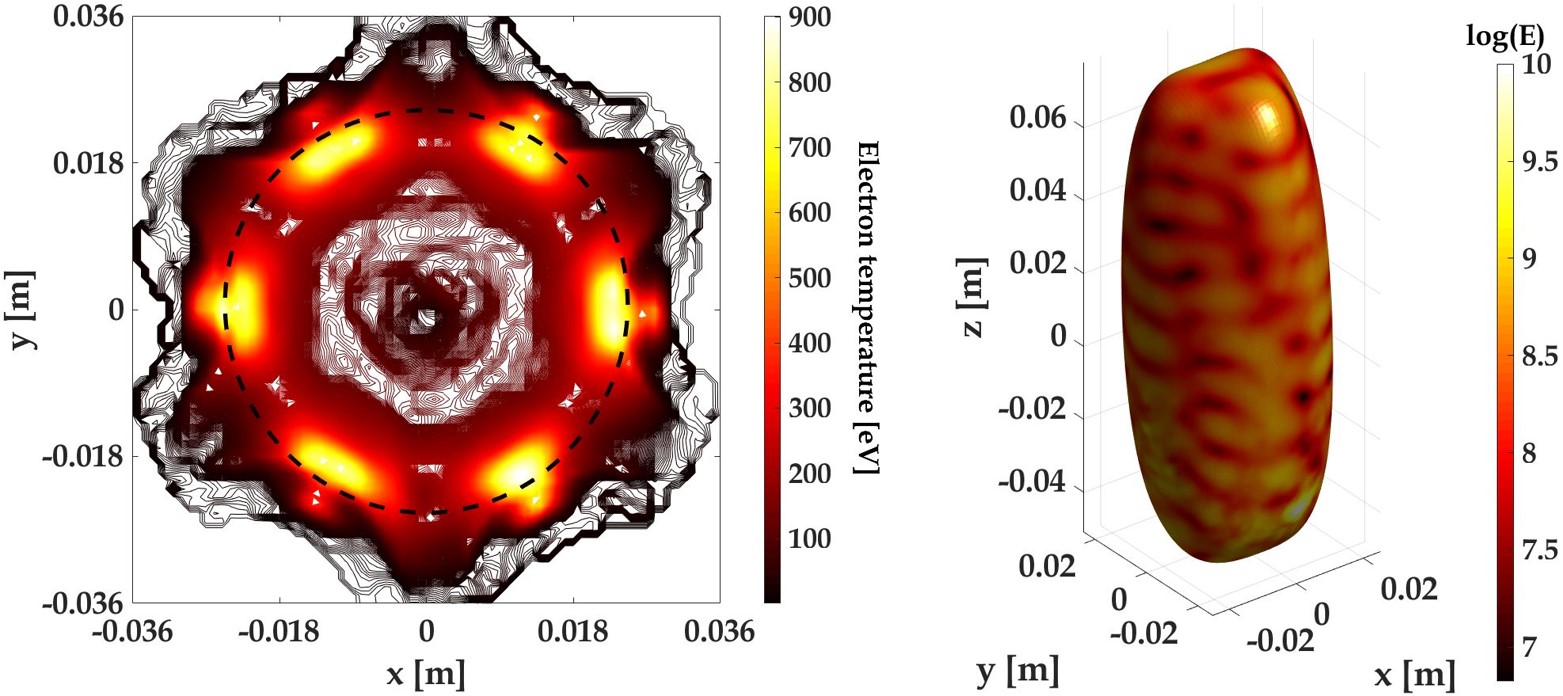}
\caption{Transversal distribution of the calculated electron temperature in the midplane (left); the black dotted line indicates the resonance. Logarithmic plot of the distribution of the electromagnetic field on the resonance surface (right). Both pictures are obtained at the step3 of the self-consistent loop.}
\label{fig:4}
\end{figure}

As mentioned in section~\ref{sec:2}, the code is able to calculate the spatial distribution of the electrons temperature: figure~\ref{fig:4} shows its transversal distribution in the midplane.
It can be seen that the maxima are located at specific points around the resonance surface (the black dotted line in the figure), outside the so-called magnetic loss cones. It is important to point out that the values in the plasmoid are everywhere lower than the one considered for the step-wise structure (1 keV), while in the halo the opposite happens in several points. By looking at the previous pictures it can be said that the density and the energetic content of the plasma are concentrated in different areas of the plasma: most of the electrons are, in fact, very well confined inside the plasmoid, while the most energetic electrons are localized in ``hot spots'' around the resonance, whose distribution is determined by the shape of the electromagnetic field (see the right part of figure~\ref{fig:4}).

\section{Charge breeding simulations}
The plasma obtained through the self-consistent loop has been used as a target to simulate the charge breeding of Rb$^{1+}$ ions, and compare the results with those described in~\cite{PSST_CB} (reference simulations from now on).  The comparison regards the reproduction of an experimental $^{85}$Rb$^{1+}$ efficiency curve as a function of the applied $\Delta V$: this parameter, equivalent to an injection energy, is determined by the difference between the high voltage $V_{1+}$ applied to the 1+ source mounted upstream the charge breeder, and the one applied to the charge breeder itself $V_{CB}$, both referred to ground. Simulations consider the $\Delta V$ as the difference between $V_{1+}$ and the plasma potential $V_P$, again both referred to ground. This means that the numerical $\Delta V$ is always smaller than the experimental one: for this reason, in order to have a correct comparison between numerical and experimental efficiency curves, the former are allowed to shift slightly toward higher values of the $\Delta V$, and such shift could be considered a measure of the plasma potential with respect to the charge breeder ($V_P-V_{CB}$). More, as a further condition for the validation of the results, the total capture at a $\Delta V$ value (including the shift) around 12 V should be higher than 40\%. The plasma consists of oxygen ions, with an average charge $\langle z\rangle$=3 and an ion temperature $KT_i$=0.3 eV, this last parameter having been deduced from the simulations. It is worth remembering that the reference simulation in~\cite{PSST_CB} was obtained with both density and temperature step-wise profiles, considering $n_{plasmoid}\sim$ 2$\cdot$10$^{17}$ m$^{-3}$, $n_{halo}$= $n_{plasmoid}$/100, $KT_e(plasmoid)$=1 keV and $KT_e(halo)$=0.1 keV. Ionizations of the injected ions have been implemented using the Lotz formula~\cite{Lotz} and a Monte Carlo approach. The comparison between charge breeding simulations with a step-wise plasma model and a self-consistent one proceeded in two phases:
\begin{enumerate}
\item First, fully step-wise vs. self-consistent density and step-wise temperature model
\item Second, fully step-wise vs. fully self-consistent model
\end{enumerate}
Both phases will be described separately in the following.
\subsection{Step-wise model vs. self-consistent density and step-wise temperature}
\begin{figure}%[ht]
\centering
\includegraphics[width=0.6\textwidth]{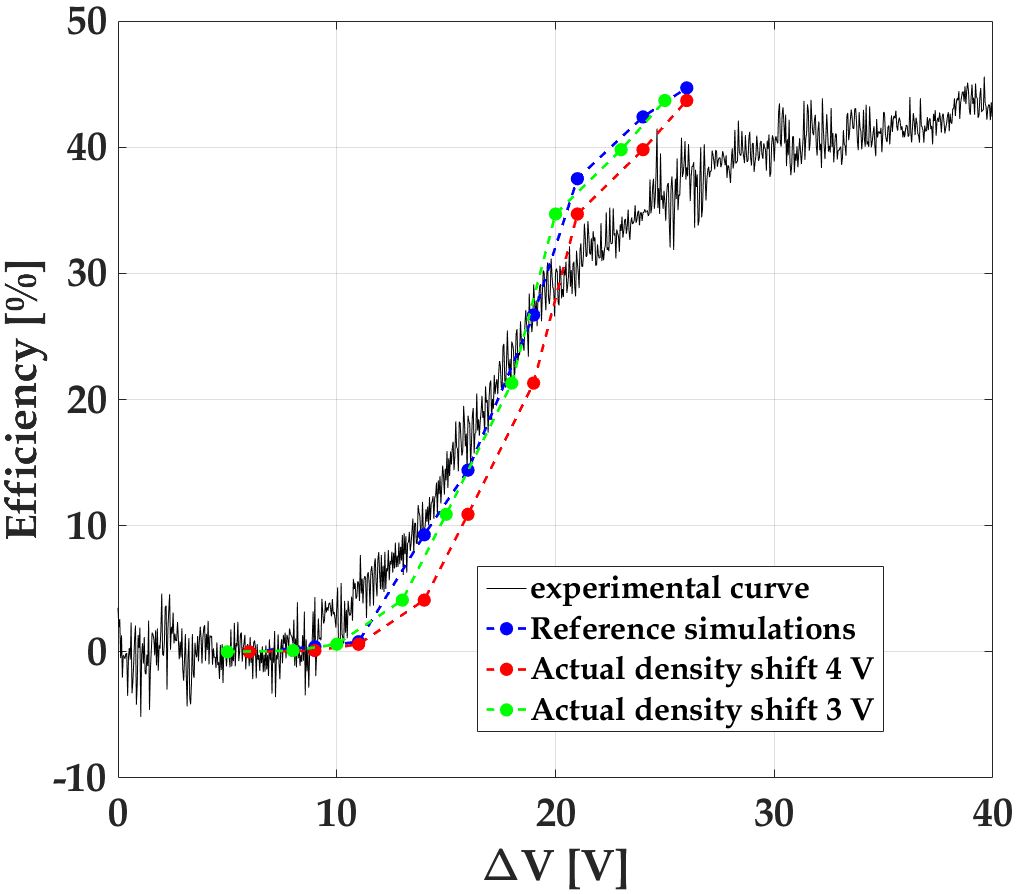}
\caption{Comparison between experimental and simulated efficiency curves of $^{85}$Rb$^{1+}$ ions for the reference simulations (blue) and those obtained considering the actual plasma density with a shift of -4 V (red) and -3 V (green).}
\label{fig:5}
\end{figure}

The results of the reference simulations have been compared to those obtained considering the self-consistent plasma density distribution, but keeping the step-wise electrons temperature: table~\ref{tab:1} shows the values of the global capture in percentage of the injected ions for different injection energies, the value in bold indicating the one corresponding to the optimum experimental $\Delta V$ as found from the reference simulations (optimum injection energy from now on). It can be seen that, besides the difference of 7\%, the global capture obtained with the self-consistent plasma density is in line with that of the reference simulations, and both with what found experimentally.
\begin{table}\footnotesize%[h]
 \centering
\caption{Comparison of the total capture of $^{85}$Rb$^{1+}$ ions injected into an oxygen plasma between the reference simulations and a plasma modified with the density profile obtained from the self-consistent loop. The values in bold indicate those corresponding to the optimum experimental $\Delta V$. }
\begin{tabular}{ccc}
\hline
&\textbf{Refernce simulations}&\textbf{Actual density, step-wise temperature}\\
\hline
\textbf{E}$_{inj}$ [eV]&\textbf{Total Capture}[\%]&\textbf{Total Capture}[\%]\\
\hline
2&36.1&39.1\\
5&44.4&54.9\\
\textbf{7}&\textbf{47.6}&\textbf{54.6}\\
10&39.5&44.9\\
12&21.7&32.9\\
15&7.1&14.9\\
17&2.9&7.3\\
20&0.9&2.1\\
22&0.2&1.4\\
\hline
\end{tabular}
\label{tab:1}
\end{table}
The comparison of the efficiency curves is shown in figure~\ref{fig:5}: for the sake of clarity, the results of the simulations using the actual plasma density have been plotted applying the same $\Delta V$ shift of the reference simulations (4 V) and another one (3 V) giving a better match with both the experimental curve and the one of the reference simulation.
 At this stage we demonstrated that the applied density scaling to the step-wise plasmoid/halo structure, used in the self-consistent loop, is able to create a plasma whose effect is equivalent to the one used in the reference simulations.
 %Anyway, a lower value of the shift applied, together with the observed decrease of the optimim injection energy, indicate that the two necessary conditions to evaluate an agreement with the experimental results are not satisfied at the same time. More, the discrepancy between experimental and numerical efficiency curves at high injection energy is still present.
\subsection{Step-wise vs. fully self-consistent model}
\begin{figure}%[ht]
\centering
\includegraphics[width=0.55\textwidth]{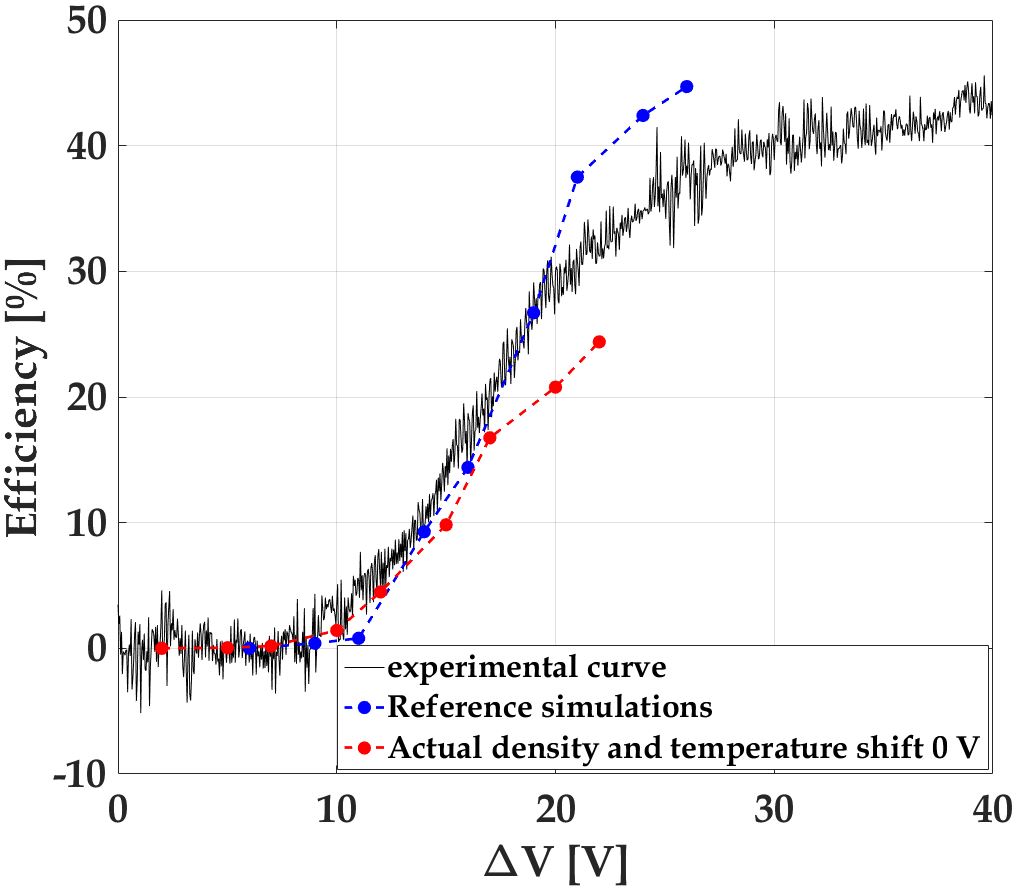}
\caption{Comparison between experimental and simulated efficiency curves of $^{85}$Rb$^{1+}$ ions for the reference simulations (blue) and those obtained considering the actual plasma density and temperature. No $\Delta V$ shift is applied in this last case.}
\label{fig:6}
\end{figure}

\begin{figure}%[ht]
\centering
\includegraphics[width=0.6\textwidth]{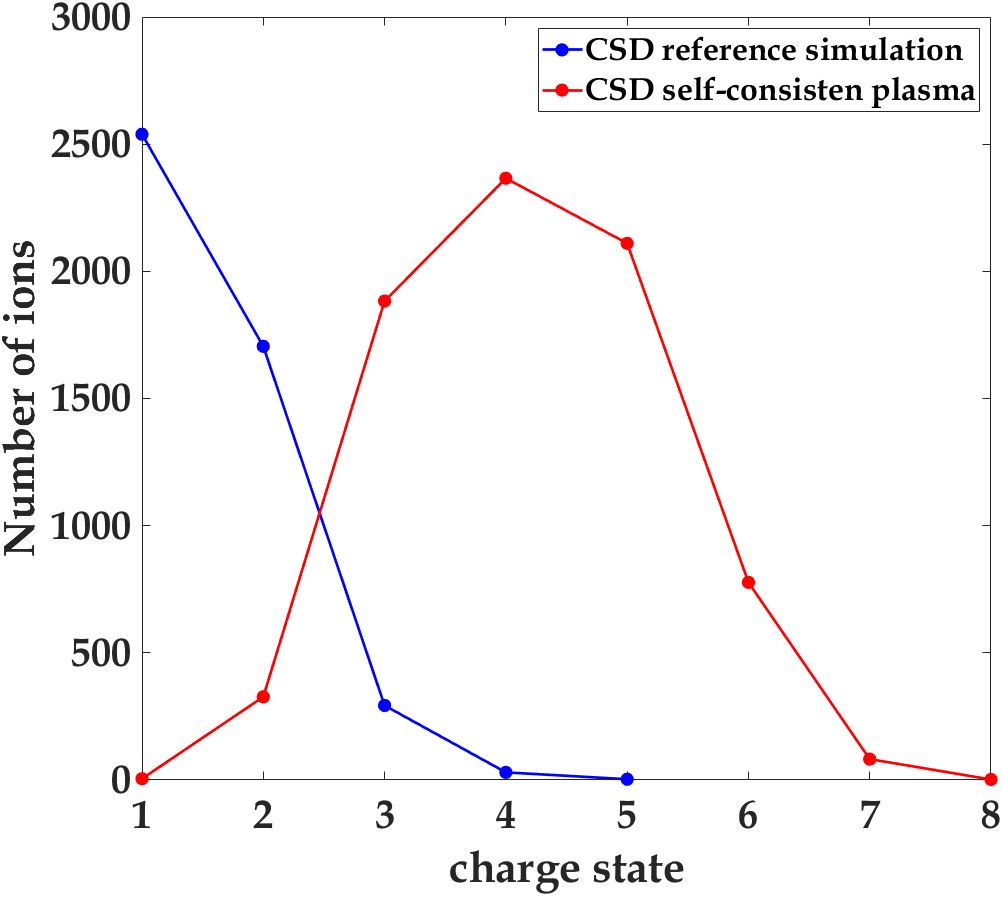}
\caption{Charge state distribution (CSD) of $^{85}$Rb ions captured by the plasma for the reference simulation (blue) and the complete self-consisten plasma (red) for an injection energy E$_{inj}$=7 eV.}
\label{fig:7}
\end{figure}

This case involved the use of the complete self-consistent plasma description, that is the actual density and temperature. Table~\ref{tab:2} shows the results for the global capture: it can be seen that, even if the maximum capture is observed at the same injection energy in both cases, the values obtained with the complete self-consistent plasma model are not only much higher than those of the reference simulations, but also than what is observed experimentally. 
\begin{table}\footnotesize%[h]
 \centering
\caption{Comparison of the total capture of $^{85}$Rb$^{1+}$ ions injected into an oxygen plasma between the reference simulations and a plasma modified with the density and temperature profiles obtained from the self-consistent loop. The values in bold indicate those corresponding to the optimum experimental injection energy E$_{inj}$. }
\begin{tabular}{ccc}
\hline
&\textbf{Refernce simulations}&\textbf{Actual density and temperature}\\
\hline
\textbf{E}$_{inj}$ [eV]&\textbf{Global Capture}[\%]&\textbf{Global Capture}[\%]\\
\hline
2&36.1&48.0\\
5&44.4&72.6\\
\textbf{7}&\textbf{47.6}&\textbf{75.0}\\
10&39.5&70.6\\
12&21.7&61.1\\
15&7.1&44.1\\
17&2.9&35.1\\
20&0.9&23.2\\
22&0.2&17.1\\
\hline
\end{tabular}
\label{tab:2}
\end{table}
The discrepancy between the step-wise and the complete self-consistent plasma model is more evident plotting the $^{85}$Rb$^{1+}$ efficiencies in figure~\ref{fig:6}: the curve obtained with the complete self-consistent plasma cannot be superimposed to the experimental one even without considering any shift in $\Delta V$.
A possible explanation for this discrepancy can be led back to the influence of the ionization process on the capture efficiency. Figure~\ref{fig:7} shows the comparison between the charge state distribution of the captured $^{85}$Rb ions for the step-wise case and the complete self-consistent plasma, at an injection energy E$_{inj}$=7 eV: as can be seen, in the latter case captured ions undergo higher ionization rate compared to the former. This effect can be thought as a direct consequence of two aspects: on one hand, the different density distribution, with higher values for the complete self-consistent plasma compared to the step-wise case (see figure~\ref{fig:3}); on the other hand, the electron temperature is not constant in the domain of the simulation (see figure~\ref{fig:4}), and assumes values everywhere lower in the plasmoid and in several locations of the halo higher than the step-wise case.  This translates into an optimization of the cross section of the first ionizations: the combination of the aforementioned effects leads to a higher Coulomb collisions frequency, thus improving the overall capture and preventing 1+ ions to escape the plasma.

\begin{figure}[ht]
\centering
\includegraphics[width=1\textwidth]{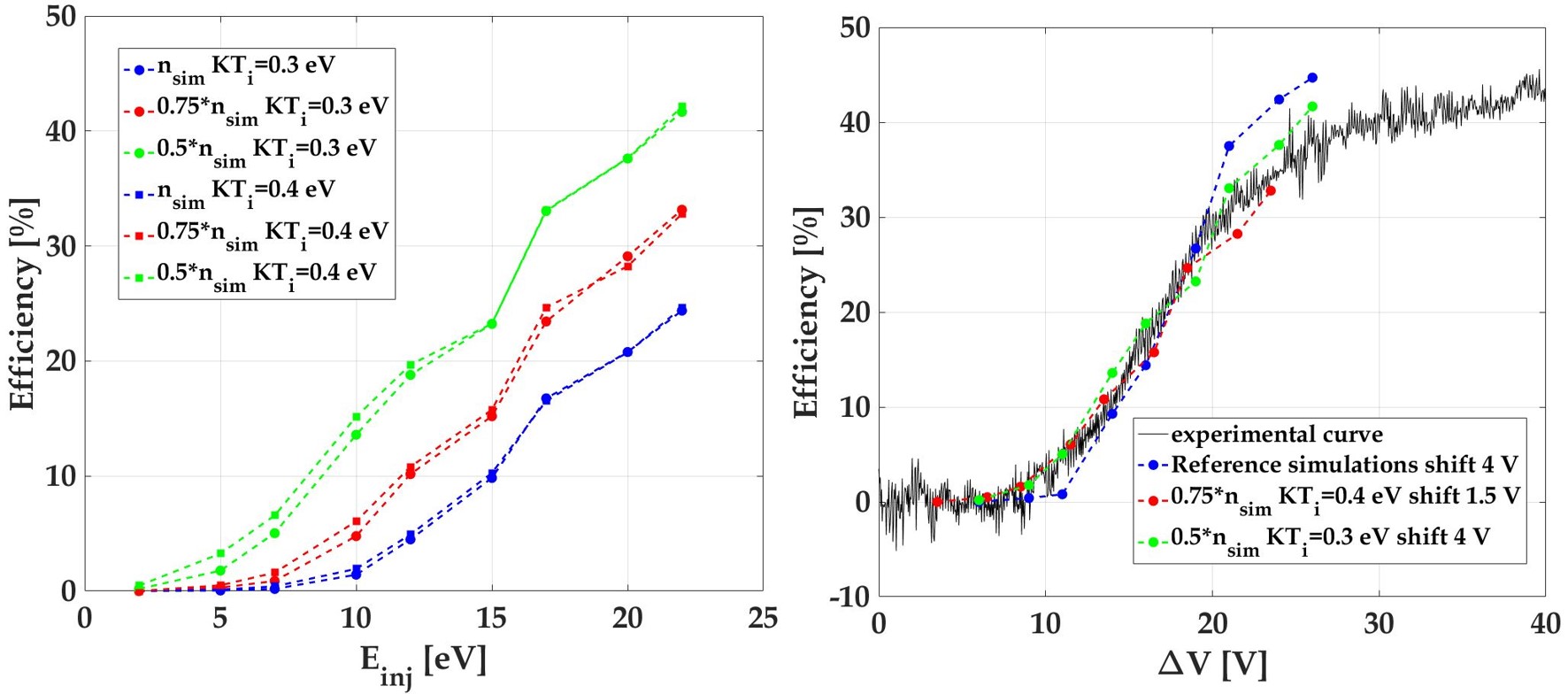}
\caption{Left: comparison of simulated efficiency curves of $^{85}$Rb$^{1+}$ ions for different plasma parameters as a function of the injection energy. Right: comparison between experimental and simulated efficiency curves of $^{85}$Rb$^{1+}$ ions for the reference simulations (blue), the case 0.75*n$_{sim}$ -- KT$_i$=0.4 eV (red) and 0.5*n$_{sim}$ -- KT$_i$=0.3 eV (green). The values of the $\Delta V$ shift applied are also shown.}
\label{fig:8}
\end{figure}

The results shown up to now indicate that the complete self-consistent plasma model leads to $^{85}$Rb$^{1+}$ efficiency curves similar to those found experimentally, that anyway cannot be properly superimposed to them, and are accompanied by a too high global capture. The analysis made with the step-wise plasmoid/halo structure of the influence of different plasma parameters on the capture process, in particular the plasma density and the ion temperature, revealed us that the extraction of 1+ ions from the charge breeder within the simulation time is possible only if the density is below about 1/10$^{th}$ of the critical density corresponding to 14.521 GHz, while a reasonable ion capture is obtained only for ion temperatures of 0.5 eV or lower~\cite{CB_Na}. The plasma obtained by the self-consistent loop leads to a too high capture and a too low rate of extracted 1+ ions: for this reason, we proceeded by  slightly modifying the values of plasma density and ion temperature, but keeping the spatial and temperature distribution obtained by the self-consistent loop. The effect of this modification on the global capture can be found in table~\ref{tab:3}, showing the results for the original plasma density and ion temperature (respectively n$_{sim}$ and $KT_i$=0.3 eV), lower plasma densities (0.75*n$_{sim}$ and 0.5*n$_{sim}$), a higher ion temperature ($KT_i$=0.4 eV) and a combination of the two (0.75*n$_{sim}$ -- $KT_i$=0.4 eV, 0.5*n$_{sim}$ -- $KT_i$=0.4 eV).
\begin{table}\footnotesize%[h]
 \centering
\caption{Comparison of the total capture of $^{85}$Rb$^{1+}$ ions injected into an oxygen plasma obtained for the complete self-consisten plasma model (n$_{sim}$ and $KT_i$=0.3 eV), lower plasma densities (0.75*n$_{sim}$ and 0.5*n$_{sim}$), a higher ion temperature ($KT_i$=0.4 eV) and a combination of the two (0.75*n$_{sim}$ -- $KT_i$=0.4 eV, 0.5*n$_{sim}$ -- $KT_i$=0.4 eV). The values in bold indicate those corresponding to the optimum experimental injection energy E$_{inj}$. }
\begin{tabular}{cccc}
\hline
&\textbf{n$_{sim}$}&\textbf{0.75*n$_{sim}$}&\textbf{0.5*n$_{sim}$}\\
&\textbf{KT$_i$=0.3 eV}&\textbf{KT$_i$=0.3 eV}&\textbf{KT$_i$=0.3 eV}\\
\hline
\textbf{E}$_{inj}$ [eV]&&\textbf{Global Capture}[\%]&\\
\hline
2&48.0&47.9&45.4\\
5&72.6&67.3&57.0\\
\textbf{7}&\textbf{75.0}&\textbf{66.6}&\textbf{50.1}\\
10&70.6&54.8&32.4\\
12&61.1&41.1&20.6\\
15&44.1&25.9&10.2\\
17&35.1&19.4&6.3\\
20&23.2&10.2&2.3\\
22&17.1&6.4&0.9\\
\hline
&\textbf{n$_{sim}$}&\textbf{0.75*n$_{sim}$}&\textbf{0.5*n$_{sim}$}\\
&\textbf{KT$_i$=0.4 eV}&\textbf{KT$_i$=0.4 eV}&\textbf{KT$_i$=0.4 eV}\\
\hline
\textbf{E}$_{inj}$ [eV]&&\textbf{Global Capture}[\%]&\\
\hline
2&38.3&36.4&33.5\\
5&54.3&47.8&39.4\\
\textbf{7}&\textbf{55.4}&\textbf{46.1}&\textbf{34.3}\\
10&49.3&36.3&22.6\\
12&42.0&27.0&13.6\\
15&29.3&16.9&6.9\\
17&24.6&12.4&4.3\\
20&15.8&7.0&1.7\\
22&11.7&4.3&0.7\\
\hline
\end{tabular}
\label{tab:3}
\end{table}
 It can be seen how lowering the density or increasing the ion temperature influences the global capture, the effect of the ion temperature being stronger. In fact, by halving the plasma density the global capture at the optimum injection energy decreases of about the 30\%, while increasing the ion temperature of the 33\% (from 0.3 eV to 0.4 eV) causes a comparable decrease of the global capture. By comparing table~\ref{tab:2} with table~\ref{tab:3}, the combination of the plasma parameters leading to the same values of the global capture as the reference simulation are 0.5*n$_{sim}$ -- $KT_i$=0.3 eV and 0.75*n$_{sim}$ --- $KT_i$=0.4 eV. The influence of the modified plasma parameters on the $^{85}$Rb$^{1+}$ efficiency curves is shown in the left part of figure~\ref{fig:8}, plotted this time as a function of the simulated injection energy (no shift): confirming the analysis carried out in~\cite{PSST_CB}, the variation of the ion temperature doesn't affect the rate at which 1+ ions are extracted from the charge breeder, while decreasing the plasma density involves the extraction of a higher number of 1+ ions and their appearance at lower injection energies. Finally, the efficiencies of the two selected combinations of plasma parameters are compared with the reference simulation and the experimental results in the right part of figure~\ref{fig:8}: it is evident that the new curves agree much better with the experimental one, even at high $\Delta V$ values whilst the reference simulation shows a small deviation. The combination 0.5*n$_{sim}$ -- $KT_i$=0.3 eV shows the same $\Delta V$ shift as the reference simulation, but the maximum of the global capture is observed at a slightly lower injection energy (see tables~\ref{tab:2} and~\ref{tab:3}). On the contrary, the combination 0.75*n$_{sim}$ -- $KT_i$=0.4 eV shows a lower $\Delta V$ shift (1.5 V) but the maximum capture is observed at the same injection energy as the reference simulation.           
\section{Discussions and Conclusions}
\begin{figure}[ht]
\centering
\includegraphics[width=0.6\textwidth]{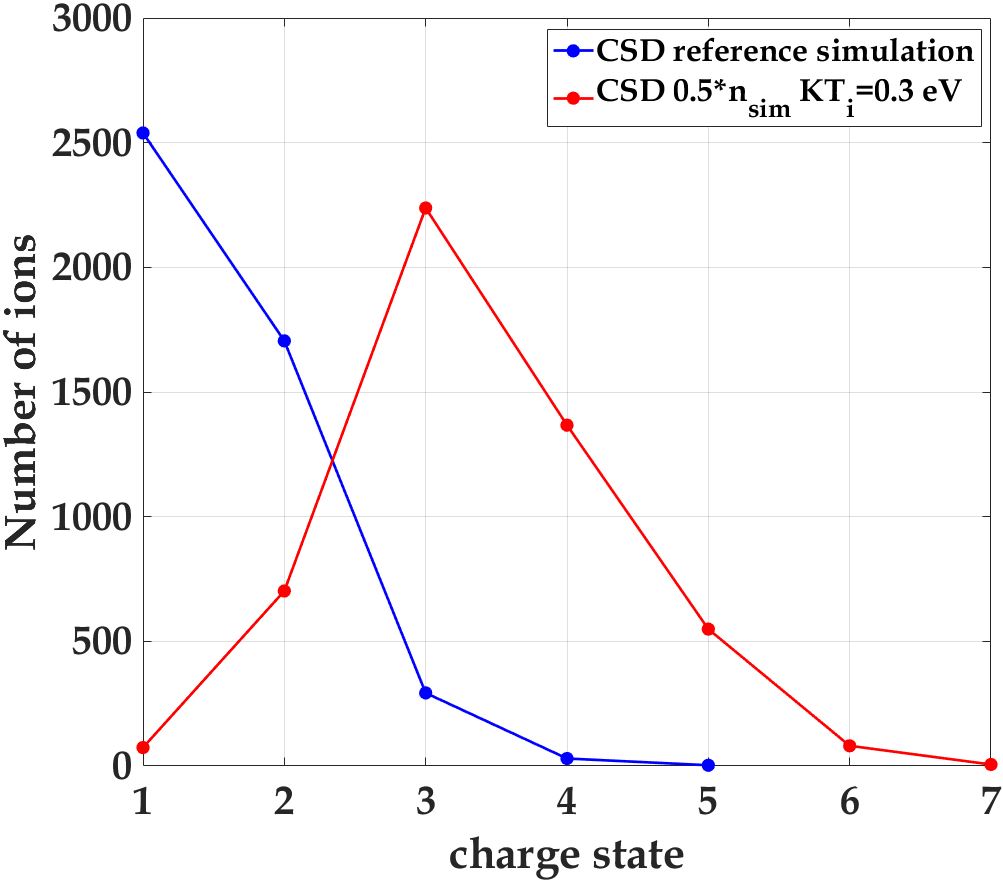}
\caption{Charge state distribution (CSD) of $^{85}$Rb ions captured by the plasma for the reference simulation (blue) and the complete self-consistent plasma with the density reduced by a factor of 2 (red) for an injection energy E$_{inj}$=7 eV.}
\label{fig:9}
\end{figure}

The simulations presented in this paper, obtained applying a self-consistent approach, showed that the plasmoid/halo structure of the ECR plasmas is automatically formed by the wave-plasma interaction for any initial distribution of plasma electrons. In fact, even though we started from initial conditions far from this structure (i.e., from uniformely distributed electrons and electromagnetic field in an empty cavity), the build-up of the plasmoid came quite soon, and it seems to converge already at the third step of the self-consistent loop. Most of the electrons are concentrated inside the resonance surface, and in this volume the density is far from being uniform: in the midplane two density peaks are observed off axis, with a decrease of around a factor of 3 going towards the middle of the plasma chamber (see figure~\ref{fig:3}). This finding confirms the possibility of a ``hollow'' nature of those plasmas, as evidenced by spatially resolved X-ray imaging. By comparing the calculated density distribution to the step-wise model used in previous publications, the maximum density is almost twice inside plasmoid, while outside the decrease is less steep than a sharp drop of two orders of magnitude. Simulations showed also the spatial distribution of the electron temperature, whose highest values are located in specific points around the resonance surface reflecting the distribution of the electromagnetic field. The implementation of the fully self-consistent plasma picture in the charge breeding simulations of $^{85}$Rb$^{1+}$ ions improved the agreement with the experimental results. Initially, the scaling with the step-wise density used in~\cite{PSST_CB} led to a higher global capture and a lower rate of the extracted 1+ ions than what observed experimentally: an explanation of this effect could be the higher density in the plasmoid and a higher cross section for the first ionizations due to the calculated temperature distribution (observable in figure~\ref{fig:7}), that joined together lead to a higher Coulomb collision frequency, thus improving the overall capture. A better agreement has then obtained by modifying slightly two plasma parameters, plasma density and ion temperature, but keeping the spatial and temperature distribution: in particular, by halving the density n$_{sim}$ obtained through the self-consistent loop, the numerical efficiency curve of $^{85}$Rb$^{1+}$  ions perfectly agrees with the experimental one, showing also the same shift in $\Delta V$ as found using the step-wise plasma structure. It is important to point out that in this case the maximum density in the plasmoid is everywhere lower than the step-wise case, but the capture capability of the plasma is the same due to the temperature distribution that optimizes the ionization process. A confirmation of this statement is given in figure~\ref{fig:9}, showing the charge state distribution of the captured ions for the step-wise case and the self-consistent case with half of the initially calculated density. It can be said that the refined calculations presented in this paper confirm three important aspects involved in the charge breeding process, in relation to what is observed experimentally and already pointed out in reference~\cite{PSST_CB}: (i) the density of the simulated charge breeder is lower of more than a factor of 10 compared to the critical density at the operating frequency, this statement being confirmed by the rate at which 1+ ions are extracted from the charge breeder; (ii) the ion temperature must be below 1 eV in order to have a global capture comparable with the experiments; (iii) the first ionizations of the injected ions play a fundamental role  in the capture process. The optimization of the developed code will continue in the next future: as a first step several temperature maps will be generated, corresponding to different electrons energy ranges, in order to be able to create an X-rays generation map directly comparable to the spectroscopic measurement described in~\cite{X-ray ATOMKI}. Then, the already available temperature map could be used to simulate the development of the charge  state distribution observed in the spectra extracted from ECR sources and charge breeders. Finally, with proper electromagnetic simulations, the effect of the injection of multiple frequencies on the electrons energy distribution could be explored. It is worth mentioning that the code seems to be now predictive, so that the effect of sources parameters will be investigated in the near future; since it seems the most relevant role is played by the electron energy distribution, scenarios where different energy distributions could be attained will be particularly explored to investigate their effect on the capture and build-up of the charge states distribution.    
%%  
%\ack
%The authors wish to thank the 5$^{th}$ Nat. Comm. of INFN, under the Grant PANDORA, for the financial support.
%%


\section*{References}
\begin{thebibliography}{99}
\bibitem{EPJ}
D. Mascali \textit{et al} 2015 \emph{Eur. Phys. J. D} {\bf 69} 27

\bibitem{ECR}
Geller R 1996 \emph{Electron cyclotron resonance ion sources and ECR plasmas} (Bristol, UK: IOP)
%
\bibitem{Pandora}
D. Mascali \textit{et al} 2017 \emph{Eur. Phys. J. A} {\bf 53} 145
%
\bibitem{SPES}
A. Galat\`a \textit{et al} 2018 \emph{JINST} {\bf13} C12009
%
\bibitem{X-ray ATOMKI}
R. Racz \textit{et al} 2017 \emph{Plasma Sources Sci. Technol} {\bf 26} 075011
%
\bibitem{Wiesemann}
 A. A. Ivanov, K. Wiesemann 2005 \emph{IEEE Trans. Pl. Sci.} {\bf33} 6
%
\bibitem{PSST_CB}
A. Galat\`a \textit{et al} 2016 \emph{Plasma Sources Sci. Technol} {\bf 25} 045007
%
\bibitem{Turco}
G. Torrisi \textit{et al} 2014 \emph{Journal of Electromagnetic Waves and Applications } {\bf28} 9
%
\bibitem{Spitzer}
	L. Spitzer Jr. 1956 \textit{Physics of Fully Ionized Gases}, Interscience, New York
%
\bibitem{SPES_CB}
A. Galat\`a \textit{et al} 2016 \emph{Rev. Sci. Instrum.} {\bf87} 02B503

\bibitem{CB_EM_SIM}
A. Galat\`a \textit{et al} 2016 \emph{Rev. Sci. Instrum.} {\bf87} 02B505

\bibitem{Lotz}
W. Lotz 1967 \emph{Zeitschrift ftir Physik} {\bf206} 205--211

\bibitem{CB_Na}
O. Tarvainen \textit{et al} 2016 \emph{Phys. Rev. Accel. Beams} {\bf19} 053402





\end{thebibliography}
\end{document}